\documentclass[11pt,letterpaper]{article}
\usepackage[top=2.25cm,bottom=2.5cm,left=2.5cm,right=2.5cm]{geometry}
\usepackage[T1]{fontenc}
\usepackage{lmodern}
\usepackage{microtype}
\usepackage[skip=6pt plus 2pt]{parskip}
\usepackage{titlesec}
\titleformat*{\section}{\Large\sffamily\bfseries}
\titleformat*{\subsection}{\large\sffamily\bfseries}
\usepackage{caption}
\usepackage{tcolorbox}
\usepackage{placeins}
\usepackage{cite}
\usepackage{amsmath,amssymb,amsfonts}
\usepackage{graphicx}
\usepackage{textcomp}
\usepackage{xcolor}
\definecolor{subject10}{HTML}{B58A00}
\definecolor{subject25}{HTML}{8B8F3D}
\definecolor{subject50}{HTML}{438B73}
\definecolor{subject75}{HTML}{268A9B}
\definecolor{subject100}{HTML}{173F70}
\colorlet{volumeLow}{subject10}
\colorlet{volumeMedium}{subject75}
\colorlet{volumeHigh}{subject100}
\DeclareRobustCommand{\subjectfraction}[1]{\textcolor{subject#1}{\textbf{#1\%}}}
\DeclareRobustCommand{\volumetier}[1]{\textcolor{volume#1}{\textbf{#1}}}
\usepackage[hyphens]{url}
\usepackage{latexml}
\usepackage{booktabs}
\usepackage{multirow}
\usepackage{enumitem}
\usepackage[colorlinks=true,allcolors=subject100]{hyperref}
\hypersetup{
  pdftitle={Scaling Laws for EEG Decoding: How Much Data Is Enough?},
  pdfauthor={Jose Mauricio, Bruna J. Lopes, Leo Burgund, Raphael Y. Camargo, Bruno Aristimunha}
}
\setlist[enumerate]{itemsep=3pt,topsep=5pt}

\newcommand{\paperabstract}{%
Deep learning has become a cornerstone of EEG-based brain decoding, with a growing number of architectures proposed every day. However, how the performance of these different models scales with data volume is not clear. Although this relationship has been characterized in other fields under the name of scaling laws, it remains poorly understood in the EEG domain. The present study addresses this gap by investigating how scan time and subject diversity affect the performance of different architectures. We evaluated five models across four EEG datasets. Training data volume was controlled by varying both subject count and trial volume under cross-subject validation. We then fitted power-law relationships to characterize the resulting behavior. Our findings reveal that as total data volume increases, the distinction between trial and subject scaling becomes largely irrelevant. Furthermore, we show that power-law relationships are both model and dataset-specific, yet they provide a robust descriptive framework for EEG decoding performance. Extrapolation to larger subject pools yields RMSE values below 0.1 in most cases. Our work contributes to the literature by providing a descriptive framework for data scaling in EEG and by demonstrating data-efficient experimental design in EEG research.}

\iflatexml
\title{Scaling Laws for EEG Decoding: How Much Data Is Enough?}
\author{Jos\'{e} Mauricio\textsuperscript{1} \and
Bruna J. Lopes\textsuperscript{2,3} \and
L\'{e}o Burgund\textsuperscript{3} \and
Raphael Y. Camargo\textsuperscript{1} \and
Bruno Aristimunha\textsuperscript{3,4}}
\date{}
\fi

\begin{document}

\iflatexml
% HTML needs flowing text instead of the PDF's fixed-size decorative box.
\maketitle
\noindent
\textsuperscript{1}\,Federal University of ABC, Santo Andr\'{e}, Brazil. \quad
\textsuperscript{2}\,Institut de Neuromodulation, GHU Paris, Paris, France. \quad
\textsuperscript{3}\,Yneuro, Paris, France. \quad
\textsuperscript{4}\,Swartz Center for Computational Neuroscience (SCCN),
Institute for Neural Computation (INC), University of California San Diego,
La Jolla, USA\par

\begin{abstract}
\paperabstract
\end{abstract}

\noindent\textbf{Keywords:} EEG decoding, scaling laws, deep learning,
brain-computer interfaces, power law\par
\noindent\textbf{Correspondence:} Bruno Aristimunha:
\href{mailto:baristimunha@ucsd.edu}{\nolinkurl{baristimunha@ucsd.edu}}\par
\else

\begin{tcolorbox}[colback=black!4,colframe=black!4,boxrule=0pt,
  arc=4mm,left=5mm,right=5mm,top=5mm,bottom=5mm,fontupper=\sffamily]
\begin{minipage}[c]{0.84\linewidth}
{\raggedright\fontsize{24}{29}\selectfont\bfseries
Scaling Laws for EEG Decoding:\\
How Much Data Is Enough?\par}
\end{minipage}\hfill
\begin{minipage}[c]{0.14\linewidth}
\raggedleft\includegraphics[height=12mm]{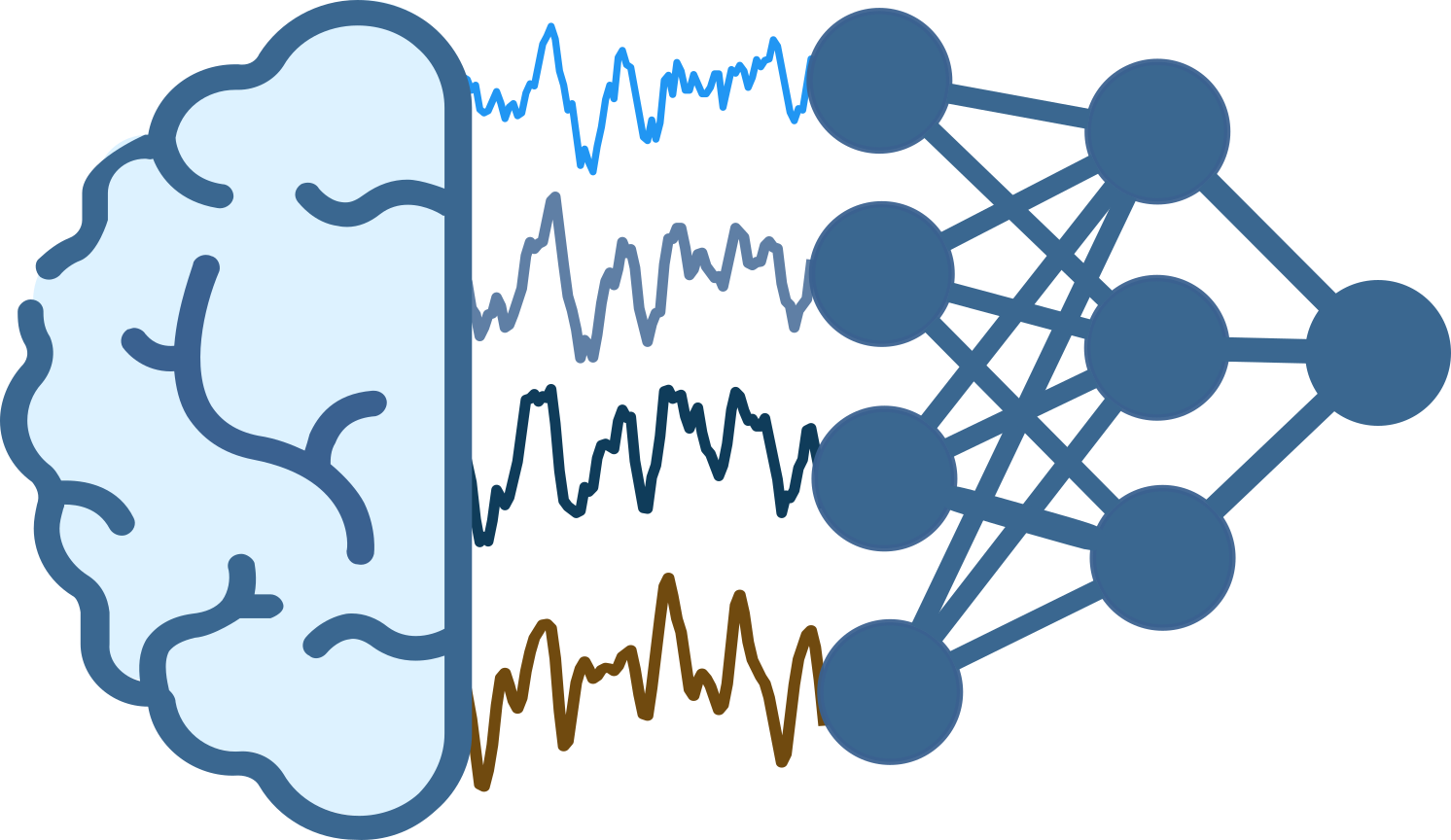}
\end{minipage}\par
\medskip

{\small\bfseries\raggedright
\mbox{Jos\'{e} Mauricio\textsuperscript{1}} \quad
\mbox{Bruna J. Lopes\textsuperscript{2,3}} \quad
\mbox{L\'{e}o Burgund\textsuperscript{3}} \quad
\mbox{Raphael Y. Camargo\textsuperscript{1}} \quad
\mbox{Bruno Aristimunha\textsuperscript{3,4}}\par}
\medskip

{\footnotesize\raggedright
\textsuperscript{1}\,Federal University of ABC, Santo Andr\'{e}, Brazil. \quad
\textsuperscript{2}\,Institut de Neuromodulation, GHU Paris, Paris, France. \quad
\textsuperscript{3}\,Yneuro, Paris, France. \quad
\textsuperscript{4}\,Swartz Center for Computational Neuroscience (SCCN),
Institute for Neural Computation (INC), University of California San Diego,
La Jolla, USA\par}
\medskip

{\small\textbf{Abstract}\par
\noindent\paperabstract
\par}\medskip
{\small\textbf{Keywords:} EEG decoding, scaling laws, deep learning,
brain-computer interfaces, power law\par}
\medskip
{\small\textbf{Correspondence:} Bruno Aristimunha:
\href{mailto:baristimunha@ucsd.edu}{\nolinkurl{baristimunha@ucsd.edu}}\par}
\end{tcolorbox}
\fi

\section{Introduction}

Electroencephalography (EEG) is the most common neurophysiological modality for decoding tasks in Brain-Computer Interfaces (BCIs) because of its non-invasiveness, low cost, and real-time recording capability. Recent advances in EEG decoding have been largely driven by Deep Learning (DL), replacing handcrafted feature engineering with end-to-end representation learning, with architectures spanning convolutions~\cite{Chen2024EEGNeX}, attention mechanisms~\cite{eegConformer}, transformers~\cite{ctNet, altaheri2022physics}, and foundation models~\cite{el2026reve}.

DL models share a heavy dependence on data volume. However, EEG data acquisition is limited by structural factors, since data can only be obtained through in-person recording sessions. Therefore, it is not only about how much training data is necessary, but also what practitioners should prioritize during data acquisition: either increasing recording time per subject or recruiting more participants.

The relationship between performance and data regime has been extensively formalized in Natural Language Processing (NLP)  \cite{hestness2017scaling, kaplan2020scaling} and Computer Vision (CV) \cite{zhai2022scaling}. The possibility of fitting empirical scaling laws relating performance to data and model has guided architectural design depending on the training and data resources available \cite{hoffmann2022training}, given that different architectures often exhibit distinct scaling behavior \cite{tay2023scaling, cao2022training}. Although this theoretical framework is well established in other fields, it remains largely unexplored in the EEG domain, despite its promising applicability.

To investigate the scaling behavior of different EEG architectures, we evaluated five different DL models across four public benchmarks with distinct task types. Models were trained under varying numbers of subjects and windowed trials, and subsequently evaluated on a fixed held-out test set of unseen subjects. Through these studies, we show that:

\begin{enumerate}
\item The performance of most tested EEG decoding models is described by power-law scaling with data volume.
\item Scaling parameters depend on the task and model architecture used.
\item Differences across volume tiers are generally larger than differences across subject and trial allocations.
\end{enumerate}

\section{Related Work}

\subsection{Scaling Laws}

Scaling laws formalize the relationship between performance and data volume. By extrapolating performance from partially converged runs, Kaplan et al. \cite{kaplan2020scaling} established a general power-law relationship of the form
\[
L(x) = c + A x^{-\alpha},
\]
where \( L(x) \) is the test loss as a function of the scaling factor \( x \), \( A \) is a task-dependent constant, \( \alpha > 0 \) determines the rate of improvement, and \( c \) is the fitted asymptotic loss floor.

Hoffmann et al.~\cite{hoffmann2022training} proposed an additive decomposition to disentangle the contributions of model size (\( N \)) and dataset size \( D \):
\[
L(N, D) = E + A N^{-\alpha} + B D^{-\beta}.
\]

\noindent Similar scaling laws have been investigated in computer vision \cite{zhai2022scaling}, graph-based models \cite{liu2024towards}, and time-series forecasting \cite{edwards2024scaling}. This raises the natural question of whether analogous scaling behaviors can be identified in EEG decoding.

\subsection{Scaling Laws with Brain Imaging Data}

Literature on scaling brain decoding models is relatively new. Ooi et al.~\cite{ooi2025longer} systematically examined how functional magnetic resonance imaging (fMRI) model performance for phenotypical prediction scales with both subject diversity and scan time. They showed that increasing data size improves model performance. Conversely, a study on image decoding from brain signals \cite{banville2025scaling}, which included both EEG and fMRI, suggests that data quantity per subject may play a more important role. However, their work was focused exclusively on image decoding.

Bomatter and Gouk \cite{bomatter2025limited} were among the first to directly explore how data affects model performance in EEG tasks. They concluded that increasing the number of participants yields greater performance gains than extending recording time per subject. Their multilevel data model describes constraints on scaling across participants. Here, we evaluate extrapolation to larger subject pools across five architectures and four datasets.

\section{Methodology}

\subsection{Datasets} Four public EEG benchmarks were used: NMT Scalp EEG Dataset \cite{khan2022nmt}, with ~1,500 participants and Temple University Abnormal EEG Corpus (TUAB) \cite{lopez2017automated} comprising 3,000 subjects, are datasets with a binary normality task. PhysioNetMI \cite{goldberger2000physiobank} dataset, involving 109 volunteers, targets motor imagery and lastly, ISRUC-Sleep dataset \cite{khalighi2016isruc}, with 118 individuals, is applied to sleep stage classification. Data were accessed through OpenEEGBench~\cite{guetschel2026openeegbench} and EEGDash~\cite{aristimunha2026eegdash}.

\subsection{Models and Hyperparameters} We evaluated five architectures (EEGNeX \cite{Chen2024EEGNeX}, CTNet \cite{ctNet}, ATCNet \cite{altaheri2022physics}, MSVTNet \cite{msvt2024}, REVE \cite{el2026reve}). All models were trained for 100 epochs using cross-entropy loss with balanced class weighting. The experimental pipeline uses Braindecode~\cite{aristimunha2025braindecode} for model construction and data handling.

\subsection{Experimental Scaling Design}
We adopted an 80/20 cross-subject split, ensuring that no subject appeared in both the training and test sets. Inspired by prior evaluations of continual fine-tuning~\cite{wimpff2025fine} and Euclidean alignment across subjects~\cite{junqueira2024alignment}, we independently varied the number of training trials (10--100\% in 10\% increments, using class-stratified sampling) and the proportion of training subjects (\subjectfraction{10}, \subjectfraction{25}, \subjectfraction{50}, \subjectfraction{75}, and \subjectfraction{100} of the available training subjects).

Each experimental condition was repeated five times using different random seeds. For each random seed, the test set was defined and kept unchanged across all scaling conditions. After defining each training partition, approximately 15\% of its trials were randomly selected to form the validation set, with the remaining trials used for model training. The scaling analysis was performed using the test loss recorded at the epoch achieving the best validation performance, with losses averaged across the five random seeds.

To isolate the effect of data composition, we conducted a secondary experiment using only the CTNet architecture, systematically varying how training data were allocated between subjects and trials. For each dataset, we defined three volume tiers (\volumetier{Low}, \volumetier{Medium}, \volumetier{High}) and, within each, three configurations: \textbf{Trial Heavy}, \textbf{Subject Heavy}, and \textbf{Balanced}. Corresponding ratios are presented in Table~\ref{tab:composition}. Realized totals could differ slightly. Due to smaller dataset sizes, the \volumetier{High} tier was restricted: only \textbf{Balanced} for PhysioNetMI, and only \textbf{Subject Heavy} and \textbf{Balanced} for ISRUC.

\suppressfloats[t]
\begin{table}[t]
\centering
\caption{Subject $\times$ Window allocations for each nominal budget.}
\label{tab:composition}
\small
\setlength{\tabcolsep}{6pt}
\begin{tabular}{llccc}
\toprule
Dataset & Budget & Window Heavy & Balanced & Subject Heavy \\
\midrule
PhysioNet
& \volumetier{Low}
& $5{\times}72$ & $12{\times}30$ & $36{\times}10$ \\
& \volumetier{Medium}
& $25{\times}72$ & $45{\times}40$ & $60{\times}30$ \\
& \volumetier{High}
& -- & $87{\times}62$ & -- \\
\midrule
ISRUC
& \volumetier{Low}
& $2{\times}700$ & $7{\times}200$ & $14{\times}100$ \\
& \volumetier{Medium}
& $10{\times}700$ & $20{\times}350$ & $35{\times}200$ \\
& \volumetier{High}
& -- & $50{\times}700$ & $70{\times}500$ \\
\midrule
TUAB
& \volumetier{Low}
& $100{\times}10$ & $200{\times}5$ & $500{\times}2$ \\
& \volumetier{Medium}
& $500{\times}10$ & $1{,}000{\times}5$ & $1{,}250{\times}4$ \\
& \volumetier{High}
& $1{,}000{\times}10$ & $1{,}250{\times}8$ & $1{,}667{\times}6$ \\
\midrule
NMT
& \volumetier{Low}
& $15{\times}130$ & $40{\times}50$ & $100{\times}20$ \\
& \volumetier{Medium}
& $154{\times}130$ & $400{\times}50$ & $1{,}000{\times}20$ \\
& \volumetier{High}
& $769{\times}130$ & $1{,}000{\times}100$ & $1{,}933{\times}52$ \\
\bottomrule
\end{tabular}
\end{table}

\FloatBarrier
\subsection{Scaling Curve Fitting}

Generalizing the framework of Hoffmann et al.~\cite{hoffmann2022training} to EEG decoding, we consider two distinct scaling factors: the number of windowed trials per subject (\(N_t\)) and the number of subjects (\(N_s\)). For each model and dataset, we fitted a generalized power-law relationship defined as
\begin{equation}
\begin{aligned}
L
={}&
A\left(\frac{N_t}{\bar{N}_t}\right)^{-\alpha}
+
B\left(\frac{N_s}{\bar{N}_s}\right)^{-\beta} + L_\infty,
\end{aligned}
\label{scaling_eq}
\end{equation}
where \(L\) denotes the test-set loss. The parameters \(A\) and \(B\) are the scaling amplitudes associated with the number of trials per subject and the number of subjects, respectively. The parameters \(\alpha>0\) and \(\beta>0\) are the corresponding scaling exponents. The terms \(\bar{N}_t\) and \(\bar{N}_s\) denote the respective median values across the training conditions and are used to normalize the scaling variables, so that the amplitudes \(A\) and \(B\) refer to the median conditions. The term $L_\infty$ is the fitted asymptotic loss under this model.

We estimated parameters by minimizing the sum of Huber losses on log residuals $\log\hat{L}-\log L$ ($\delta=10^{-3}$), using multiple initializations. Fits use \subjectfraction{10}, \subjectfraction{25} and \subjectfraction{50} of available subjects. Predictive error is measured by root mean squared error (RMSE) in the original loss units on held-out \subjectfraction{75}, and \subjectfraction{100} subject conditions.

\FloatBarrier
\section{Results}

\begin{figure}[!htbp]
\centering
\includegraphics[width=\linewidth]{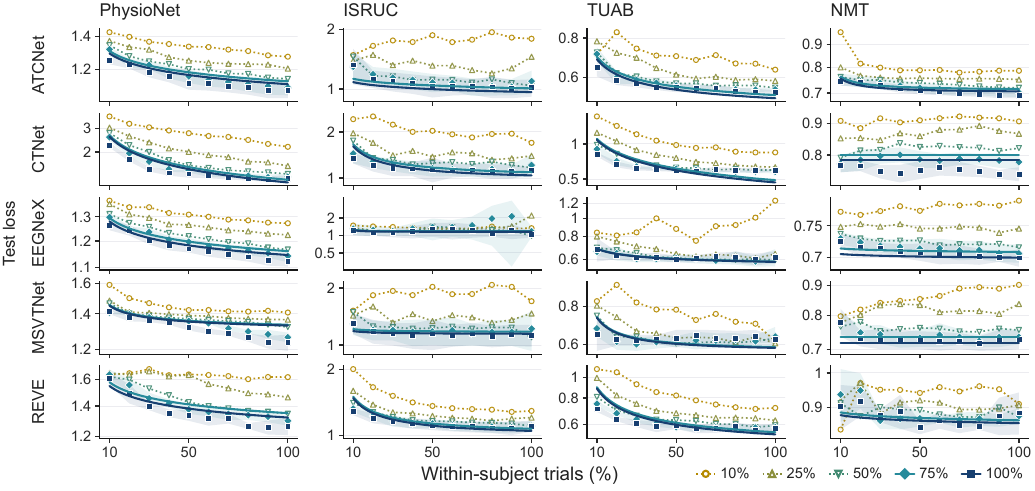}
\caption{\textbf{Scaling-law predictions at held-out subject fractions.} Columns show datasets and rows show architectures. Colours indicate subject fractions. Open markers and dotted lines show fit data (\subjectfraction{10}, \subjectfraction{25}, \subjectfraction{50}). Filled markers show held-out observations (\subjectfraction{75}, \subjectfraction{100}) and solid curves their predictions. Shading: $\pm1$ seed SD. Independent logarithmic loss scales.}
\label{fig:scaling_curves}
\end{figure}

Fig.~\ref{fig:scaling_curves} shows test set loss across data scaling regimes, revealing dataset-specific saturation thresholds. For most models, once a sufficient number of subjects is available, additional per-subject recording yields negligible returns. PhysioNet saturates at $\sim$65 subjects (\subjectfraction{75}) and $\sim$70\% of trials ($\sim$2 min). ISRUC reaches $>$85\% of achievable gain with $\sim$35 subjects (\subjectfraction{50}) and 50\% of trials (2--3 h), suggesting one night of recording suffices. For clinical EEG abnormality detection, TUAB and NMT ($\sim$930 and $\sim$967 subjects, \subjectfraction{50}) reach 90--92\% of maximum gain using only 40\% of trials ($\sim$1 min and $\sim$4 min, respectively). TUAB provides only 2.4 min of usable EEG, so its saturation may reflect sparse within-subject sampling rather than a true duration threshold. Overall, subject diversity is the binding constraint.

As this behavior closely resembles the scaling laws of Kaplan et al.~\cite{kaplan2020scaling}, we systematically fitted power-law relationships to the data. Curves are shown in Fig.~\ref{fig:scaling_curves}. Parameters and prediction errors are in Table~\ref{tab:scaling_params}.

\begin{figure}[!htbp]
\centering
\includegraphics[width=0.80\linewidth]{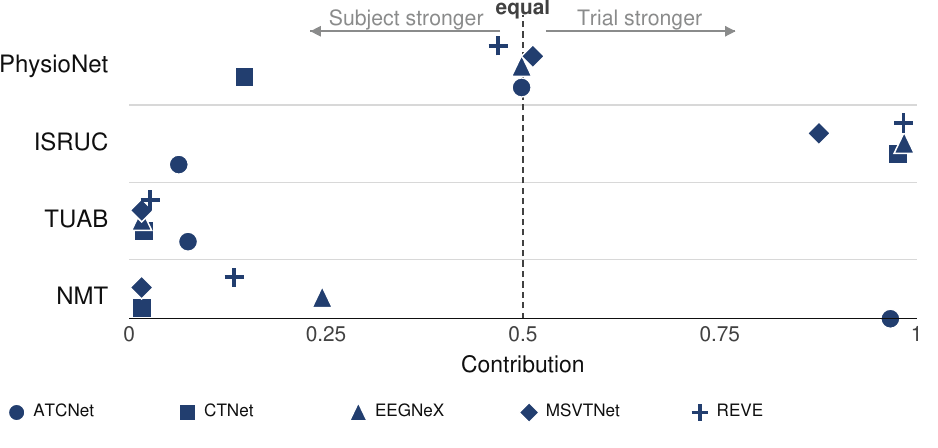}
\caption{\textbf{Relative contributions of trials and subjects.} Based on the scaling parameters from Table~\ref{tab:scaling_params}.}
\label{fig:relative_contribution}
\end{figure}

Using the fitted scaling parameters, we defined a relative scaling contribution measure (coefficients weighted to form this plot dimension). It ranges from 0 to 1: values near 1 indicate trials-driven loss, near 0 subjects-driven loss, and ~0.5 equal contributions. Results for each scaling law in Table~\ref{tab:scaling_params} are shown in Fig.~\ref{fig:relative_contribution}.

\begin{table}[!htb]
\centering
\small
\caption{Model--task fit parameters. Training-set medians shown below datasets. RMSE calculated over hold-out set.}
\label{tab:scaling_params}
\setlength{\tabcolsep}{6pt}
\begin{tabular}{llrrrrrr}
\toprule
\textbf{Dataset} & \textbf{Model} & $A_{N_t}$ & $\alpha_{N_t}$ & $B_{N_s}$ & $\beta_{N_s}$ & $L_\infty$ & RMSE \\
\midrule
\multirow{5}{*}{\shortstack[l]{PhysioNet \\ $\bar{N}_t=40$ \\ $\bar{N}_s=22$}}
& ATCNet & 1.30 & 0.07 & 0.65 & 0.31 & 0.00 & 0.03 \\
& CTNet & 5.15 & 0.39 & 9.45 & 0.87 & 0.00 & 0.15 \\
& EEGNeX & 0.78 & 0.11 & 0.92 & 0.08 & 0.00 & 0.01 \\
& MSVTNet & 0.47 & 0.49 & 0.40 & 0.62 & 1.25 & 0.05 \\
& REVE & 1.09 & 0.14 & 1.35 & 0.14 & 0.00 & 0.05 \\
\midrule
\multirow{5}{*}{\shortstack[l]{ISRUC \\ $\bar{N}_t=540$ \\ $\bar{N}_s=18$ }}
& ATCNet & 1.18 & 0.06 & 4.08 & 0.73 & 0.00 & 0.08 \\
& CTNet & 24.03 & 0.80 & 4.24 & 0.67 & 0.69 & 0.09 \\
& EEGNeX & 4.7k & 2.47 & 1.56 & 0.07 & 0.00 & 0.30 \\
& MSVTNet & 2.12 & 0.88 & 4.88 & 0.91 & 1.04 & 0.04 \\
& REVE & 17.68 & 0.80 & 0.93 & 0.27 & 0.68 & 0.07 \\
\midrule
\multirow{5}{*}{\shortstack[l]{TUAB \\ $\bar{N}_t=8$ \\ $\bar{N}_s=466$}}
& ATCNet & 0.30 & 0.49 & 1.59 & 0.34 & 0.31 & 0.02 \\
& CTNet & 1.14 & 0.43 & 16.19 & 0.67 & 0.00 & 0.11 \\
& EEGNeX & 0.19 & 0.45 & 10.4k & 1.99 & 0.52 & 0.02 \\
& MSVTNet & 0.24 & 0.92 & 58.75 & 1.16 & 0.55 & 0.04 \\
& REVE & 0.57 & 0.61 & 5.37 & 0.57 & 0.36 & 0.06 \\
\midrule
\multirow{5}{*}{\shortstack[l]{NMT \\ $\bar{N}_t=93$ \\ $\bar{N}_s=483$}}
& ATCNet & 0.87 & 0.99 & 0.60 & 0.14 & 0.49 & 0.01 \\
& CTNet & 0.00 & 0.00 & 1.31 & 0.07 & 0.00 & 0.02 \\
& EEGNeX & 0.55 & 0.00 & 0.62 & 0.17 & 0.00 & 0.01 \\
& MSVTNet & 0.00 & 0.00 & 1.36 & 0.08 & 0.00 & 0.02 \\
& REVE & 0.83 & 0.01 & 0.87 & 0.32 & 0.00 & 0.03 \\
\bottomrule
\end{tabular}
\end{table}

\begin{samepage}
In PhysioNet, neither scaling dimension consistently dominated across models. CTNet was the exception, with a greater contribution from subjects. In ISRUC, with the exception of ATCNet, most models showed a greater contribution from the Trial dimension. Notably, EEGNeX exhibited no apparent scaling with data volume on this dataset. In contrast, for NMT and TUAB, the scaling equations primarily relied on the Subject dimension. Most model and dataset pairs achieved an RMSE below 0.1 on the held-out set, indicating a close agreement between the fitted scaling functions and the observed test losses.
\par
\end{samepage}

\begin{figure}[!htbp]
\centering
\includegraphics[width=0.64\linewidth]{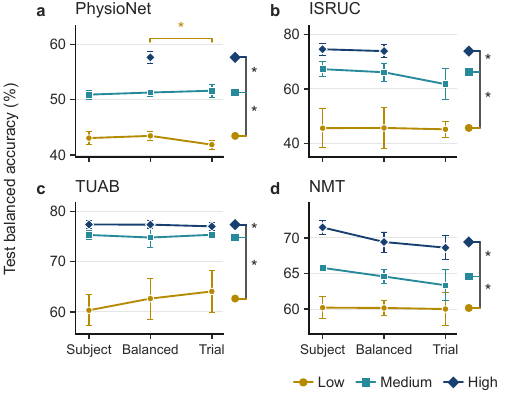}
\caption{\textbf{CTNet at matched volume.} \volumetier{Low}/\volumetier{Medium}/\volumetier{High}: window budgets. Subject/Trial: \mbox{subject-/trial-heavy}. Y-axes: mean test balanced accuracy $\pm$ SD. Brackets: allocations within tiers (horizontal) or adjacent tiers at Balanced (right). $*$: unpaired two-sided Mann--Whitney U, $p<0.05$.}
\label{fig:Matched_Dataset}
\end{figure}

\FloatBarrier
To further investigate the effect of data composition in real decoding performance, we compared different sampling allocations at matched training volumes. Larger nominal training volumes generally resulted in higher recorded peak test balanced accuracy (Fig.~\ref{fig:Matched_Dataset}). Within the same volume tiers, differences between sampling compositions were generally limited. The only notable exception was PhysioNetMI, which showed lower performance for the \textbf{Trial Heavy} composition than for the \textbf{Balanced} composition in the \volumetier{Low} volume regime.

\FloatBarrier
\section{Discussion and Limitations}

The results of this study suggest that EEG decoding performance with deep learning scales with data and that power-law relationships offer a useful descriptive framework within the tested conditions. This phenomenon has been observed across domains ranging from CV to NLP \cite{kaplan2020scaling, hoffmann2022training, zhai2022scaling, liu2024towards, edwards2024scaling}, yet it does not translate universally: in single-cell biology, Denadel et al.~\cite{denadel2026evaluating} found that performance in single-cell foundation models tended to plateau as pretraining data volume increased.

Our findings align with prior EEG work~\cite{ooi2025longer, bomatter2025limited} and further show that scaling properties are specific to both the model and the task. For instance, REVE underperformed most models at 10\% and 25\% subject volumes on PhysioNet, but rapidly became competitive as data volume increased. This suggests that small datasets such as BCI Competition IV \cite{tangermann2012review} may not be ideal for comparing EEG DL architectures, since conclusions drawn at low data volumes may not hold at scale.

More broadly, given sufficient data volume, the Trial and Subject dimensions no longer differ in performance. Practitioners aiming to optimize data acquisition should therefore prioritize subject diversity: after a large number of subjects is reached, minimal per-subject recording may suffice: $\sim$2 min of pure motor imagery (around 40 MI trials), as suggested by PhysioNet results, $\sim$3 h (a single full night of sleep) for sleep stage classification, as in ISRUC, and $\sim$4 min for abnormality detection, as in NMT.

Another important implication of our work concerns architectural design. The seminal studies of Kaplan et al.~\cite{kaplan2020scaling} and Hoffmann et al.~\cite{hoffmann2022training} are influential not only for their characterization of predictive performance in large language models, but also for the practical design heuristics they provide regarding model parametrization. In computer vision, following the introduction of DINOv1 \cite{caron2021emerging}, its successors DINOv2 \cite{oquab2023dinov2} and DINOv3 \cite{simeoni2025dinov3} examined the relationship between data scale and parameterization, informing the choice of model components. This highlights an important aspect that remains to be isolated: the role of model parameterization within a given architecture. Although the scaling analysis was repeated across multiple EEG decoding architectures\cite{eegConformer, altaheri2022physics, el2026reve}, their architectural heterogeneity makes it difficult to disentangle the effect of model size from architectural design choices.

Our findings have practical implications for data acquisition, model distillation and foundation model training. Understanding how different architectures scale with data volume can help guide decisions about which models to distill and how much data is required to train effective foundation models for EEG. Future work should systematically investigate these dimensions to develop a more comprehensive theory of scaling in the EEG domain.

\FloatBarrier
\section*{Acknowledgments}

We acknowledge computing resources provided by the San Diego Supercomputer Center: Voyager~\cite{sdsc2025voyager} (NSF award 2005369) and Expanse~\cite{sdsc2025expanse} (NSF award 1928224), accessed through allocation CIS261538 from the Advanced Cyberinfrastructure Coordination Ecosystem: Services \& Support (ACCESS) program~\cite{boerner2023access}. ACCESS is supported by U.S. National Science Foundation grants 2138259, 2138286, 2138307, 2137603, and 2138296. We also thank GENCI--IDRIS for access to the Jean Zay supercomputer under grant 2025-AD011014817R1.

{\small
\bibliographystyle{IEEEtran}
\bibliography{references}
}

\end{document}